\documentclass{article}
\pdfoutput=1
\usepackage{microtype}
\usepackage{graphicx}
\usepackage{subfigure}
\usepackage{booktabs} 

\usepackage{hyperref}

\usepackage[accepted]{icml2024}

\usepackage{amsmath}
\usepackage{amssymb}
\usepackage{mathtools}
\usepackage{amsthm}

\usepackage[capitalize,noabbrev]{cleveref}

\theoremstyle{plain}

\newtheorem{position}{Position}

\theoremstyle{definition}

\theoremstyle{remark}

\usepackage[textsize=tiny]{todonotes}

\usepackage{tikz} 
\usetikzlibrary{shapes,decorations,arrows,calc,arrows.meta,fit,positioning}
\tikzset{
    -Latex,auto,node distance =1 cm and 1 cm,semithick,
    state/.style ={ellipse, draw, minimum width = 0.7 cm},
    point/.style = {circle, draw, inner sep=0.04cm,fill,node contents={}},
    bidirected/.style={Latex-Latex,dashed},
    el/.style = {inner sep=2pt, align=left, sloped},
    pdp/.style = {pdpcolor, dash pattern = on 2pt off 4pt, dash phase=3pt, line width=0.8pt},
    tdp/.style = {tdpcolor, dash pattern = on 2pt off 4pt, line width=0.8pt},
    itdp/.style = {tdpcolor, dash pattern = on 2pt off 1pt, line width=0.8pt},
    ecmtdp/.style = {tdpcolor}
}

\icmltitlerunning{Causality and Scientific Pragmatism}

\begin{document}

\twocolumn[
\icmltitle{Position: The Causal Revolution Needs Scientific Pragmatism}



\icmlsetsymbol{equal}{*}

\begin{icmlauthorlist}
\icmlauthor{Joshua R. Loftus}{LSE}
\end{icmlauthorlist}

\icmlaffiliation{LSE}{Department of Statistics, London School of Economics, London, UK}

\icmlcorrespondingauthor{Joshua Loftus}{j.r.loftus@lse.ac.uk}

\icmlkeywords{Causality, causal inference, scientific method, pragmatism, value pluralism}

\vskip 0.3in
]



\printAffiliationsAndNotice{}  

\begin{abstract}
    Causal models and methods have great promise, but their progress has been stalled. Proposals using causality get squeezed between two opposing worldviews. Scientific perfectionism--an insistence on only using ``correct'' models--slows the adoption of causal methods in knowledge generating applications. Pushing in the opposite direction, the academic discipline of computer science prefers algorithms with no or few assumptions, and technologies based on automation and scalability are often selected for economic and business applications. We argue that these system-centric inductive biases should be replaced with a human-centric philosophy we refer to as scientific pragmatism. The machine learning community must strike the right balance to make space for the causal revolution to prosper.
\end{abstract}

\section{Introduction} 
\label{sec:intro}

There is a causal revolution underway in empirical sciences. From its century-old roots in controlled experiments, in recent decades it has made inroads in observational health and social sciences. As a research topic among methodologists, causal models have gained some popularity in economics, computer science, and statistics, and may be spreading into applications in other research fields and industries. However, \textit{there is a risk that the causal revolution will stall}, and that humanity will miss out on the potential benefits of its success. To realize these benefits, \textbf{we argue that a philosophy of scientific pragmatism is necessary for the causal revolution to flourish}. This philosophy views causal models pragmatically as tools for hypothetical reasoning.

For the sake of simplicity we discuss only two types of models: causal and predictive. Strictly speaking, causal models are a subset of predictive models, so here we will use ``predictive'' as shorthand for ``predictive models which are not also causal.'' If we think of modeling as a process of making assumptions and drawing conclusions then \emph{causal models make more assumptions than predictive models}. These additional assumptions afford causal models additional conclusions, but also open them up to more criticism.

\begin{figure}[ht]
\centering
\begin{tikzpicture}
    \node[state] (x) at (0,0) {$X$};
    \node[state] (y) [below = of x] {$Y$};
    \path (x) edge (y);
    \node[state] (x2) [right = of x] {$X$};
    \node[state] (y2) [below = of x2] {$Y$};
    \path (y2) edge (x2);
    \node[state] (x3) [right = of x2] {$X$};
    \node[state] (z) [below right = of x3] {$Z$};
    \node[state] (y3) [below = of x3] {$Y$};
    \path (z) edge (x3);
    \path (z) edge (y3);
\end{tikzpicture}
\caption{Structural causal models (SCMs) represented as directed acyclic graphs (DAGs). Variables are the nodes of the graph and causal effects are represented by arrows. An intervention is an operation that modifies the graph in some way. Intervening on a variable means we erase the arrows pointing into that variable, set the value of that variable arbitrarily, and then propagate the new value along directed pathways pointing out of that variable. On the left, $X$ is a cause of $Y$, so intervening on $X$ will result in a change in $Y$. In the other cases, intervening on $X$ results in no change in $Y$. It is possible that the accuracy of some function $f(X)$ in predicting $Y$ is equal in all cases.}
\label{fig:direction}
\end{figure}
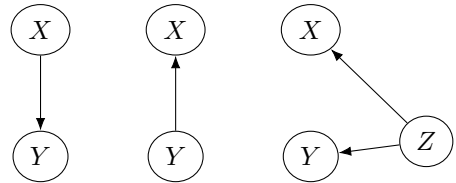

In the scenario of Figure~\ref{fig:direction}, if we only judge a model based on its predictions of some important variable $Y$ then the same model may work equally well regardless of the direction of causality. If we expect a model to also tell us what happens when we intervene on variables, then these models with equal predictive accuracy are no longer equivalent. \emph{We become responsible to choose one of them}.

We can use causal models to answer important questions that predictive models cannot illuminate. For example, if we want to know how we might change $Y$, then the model on the left of Figure~\ref{fig:direction} says we can intervene on $X$, the model on the right says we can intervene on $Z$ (and that $X$ would also change in that case), and the model in the middle says we would have to intervene on $Y$ directly (which changes $X$ as well).

The differences between these cases have important consequences when we apply models like these to reason about the real world. Suppose that $Y$ is some variable we would like to change, but due to real world constraints we can only intervene on $X$. We have data and sophisticated predictive models that predict $Y$ from $X$ with state of the art (SOTA) accuracy. Does this mean we should then spend resources on an intervention that changes $X$? The predictive model alone cannot answer this, we have to choose a causal model. If we choose the model on the left of Figure~\ref{fig:direction}, but the real world is better described by the one on the right, then we may end up wasting resources, harming people, or pursuing a research program that will end in some type of failure. 

Causal models have more assumptions and more implications. They force us to make more choices. As a result, there are more ways our choices can go wrong, but also more potential applications. To access the benefits of a causal revolution we have to be willing to pay the costs. We must learn to strike the right balance when judging proposals that involve causal modeling.

\subsection{Published Context} 
\label{sec:published_context}

Before elaborating our positions, we provide some context about the status of the causal revolution in empirical sciences. We would like to know if scientists care about causality and whether they are modeling and writing about it in a justified way. Fortunately for our purposes, meta-scientists have been hard at work studying these questions. Some indicative findings:

\begin{itemize}
    \item
    In observational studies from 4 leading journals of obesity and nutrition (in 2006), 31\% of papers had inappropriate language implying causality in the titles or abstracts \cite{cofield2010use}.

    \item
    Among all observational studies in the top 4 medical journals in 2010, authors recommend a medical practice in 56\% (and only 14\% of these recommended an RCT should be done to confirm their observational results) \cite{prasad2013observational}.

    \item
    An NLP model applied to ``about 38,000 observational studies in PubMed'' showed about 32\% used direct causal language, and most of these instances occurred only in the conclusion sections (i.e. not in abstracts or titles) \cite{yu2019detecting}.

    \item
    A study found ``spin'' in 84\% of abstracts ``of all non-randomized studies assessing an intervention published in the BioMed Central Medical Series journals between January 1, 2011 and December 31, 2013'' and the most common kind of spin was causal language, occurring in 53\% of abstracts \cite{lazarus2015classification}.

    \item
    A study of non-RCTs in ``1,170 articles from 18 high-profile medical/public health/epidemiology journals (65 per journal) published from 2010–2019'' found ``few studies explicitly declared an interest in estimating causal effects, the majority used language that moderately or strongly implied causality'' and ``action recommendations were identified in 60.3\% [...] of discussion sections, about twice that in abstracts'' \cite{haber2022causal}.

    \item
    In a study of the 50 health studies that were most shared on social media in 2015, ``34\% percent of academic studies and 48\% of media articles used language that [human expert] reviewers considered too strong for their strength of causal inference'' \cite{haber2018causal}.

    \item
    A systematic review of 199 systematic reviews of observational studies (published in 2019) found most (57\%) were explicitly addressing causal issues, and among those which were not explicit about causal intent roughly half (51\%) used causal language \cite{han2022causal}.
\end{itemize}

The available research on ``spin'' and unjustified causal language is concentrated in health/medical sciences, but we believe the problem is much more widespread than directly indicated by the literature above. The health sciences have established traditions of systematic reviews, meta-analysis, and (pre-)registered trials. There are reasons to think the problem would only be worse in other fields where these accountability mechanisms have yet to catch on. For example, other fields have been shown to suffer from irreproducibility in their published literature \cite{open2015estimating}, and failure to properly model causality may be one of the reasons that replication studies fail.

\section{Positions} 
\label{sec:positions}

In this section we sometimes refer to the concept of an equivalent predictive model. Given any causal model, we can simply ``forget'' the additional causal assumptions regarding interventions and directionality and obtain an equivalent predictive model.

Positions~\ref{pos1}-\ref{pos4} should prove relatively uncontroversial since they have stronger logical and/or evidential support. We still find it valuable to state and argue for them because, even though they may be uncontroversial, they may still be under-appreciated. Positions~\ref{pos5}-\ref{pos6} are more speculative. We do not have evidence to cite for them. The reader will have to judge them based on their own experiences. \\

\begin{position}[Proving too much]
For any argument against using a causality in general, or a specific causal method in particular, we should check that this argument does not also apply to an equivalent predictive model.
\label{pos1}
\end{position}

This is a simple logical point but may too often be forgotten.

There are many ways that probabilistic or mathematical models in general can be criticized. Sometimes people rediscover an old, general criticism and apply it to the newest wave of research. This can be a good thing. It is worth noticing when a new modeling approach is not immune to a long-standing general problem. But we should also take care to not penalize any specific class of models based on a general critique. 

As an example, consider using a causal model to analyze fairness and discrimination, and suppose we have a variable in this model that represents a social category like race or gender. It is reasonable to use the concept of construct validity \cite{jacobs2021measurement} to criticize any model including such a variable. So it would be selective if that critique was only applied to causal models. On the other hand, it \emph{is} reasonable to criticize causal models specifically with an argument that focuses on the experimental non-manipulability of variables like race and gender \cite{sen2016race, kohler2018eddie, hu2023race, hu2020s}. And in response to good criticisms, it may be possible to adapt and improve causal models for those purposes \cite{bynum_disaggregated_2021, bynum2023counterfactuals, bynum2024new}. \\

\begin{position}[People want causality]
People usually want to reach causal conclusions. Even when they use predictive models, they often take actions or discuss the results as if the model were causal.
\label{pos2}
\end{position}

We saw evidence for this position already in the literature cited in Section~\ref{sec:published_context}. Pushing beyond the evidence, we believe this problem is more prevalent and impactful than indicated by peer reviewed scientific work. This is because many scientists are aware that ``association (e.g. correlation) is not causation,'' and hence may avoid using causal language when describing their predictive models. However, they may still be thinking about their work causally and taking actions as if the conclusions are causal. For example, they may choose to study a question that is only interesting or useful if the conclusions are causal, but they may do so using only predictive models. \\

\begin{position}[Balanced standards]
Predictive models are often used in place of causal models because they are held to too low a standard.
Conversely, causal models--and methods or papers that use them--are often held to a higher standard. This is usually wrong.
\label{pos3}
\end{position}

An author choosing to use a causal model makes additional explicit assumptions relative to an equivalent predictive model. At the stage of peer review, a referee may criticize any of these additional assumptions. And since causal models entail additional conclusions or implications, if the paper is published it is possible some subsequent research will falsify one of those additional conclusions, thereby calling the paper into question. Not wanting to risk it, scientists may avoid these costs of causality and opt for a model with fewer assumptions.

The principle of parsimony (or simplicity) states ``\emph{other things being equal}, simpler theories are better [emphasis added]'' \cite{sep-simplicity}. Now, it is true that a causal model has more assumptions than an equivalent predictive model, so in a sense the predictive model is simpler. But it is \emph{not} true that these two models are otherwise equal. The causal version implies more.

There are some good reasons to apply different standards to causal and predictive models. If an assumption of a causal model is difficult or impossible to test via any experiment, we should treat that assumption with caution and transparently acknowledge the limitation of any conclusions based on that model. All further conclusions or models built on that assumption must never forget the untested part of their foundation. Similarly, when reporting results of a predictive model we must acknowledge they do not necessarily imply any causal relationships.

Achieving a level playing field between causal and predictive models requires applying balanced standards. If a research project proposes using predictive models it should be questioned whether the desired goals of the inquiry only include predictive conclusions. If the work does not pay the price of making explicit assumptions required for causal conclusions it should not be allowed to suggest those conclusions for free. On the other hand, if causal conclusions are desired, then it is wrong to penalize the additional assumptions necessary for the model to able to answer our actual questions. \\

\begin{position}[Falsifiability and the scientific method]
Causal models can be falsified in more ways than predictive models. This is usually good.
\label{pos4}
\end{position}

Falsificationism, one of the most well-known philosophies about the scientific method, states that science proceeds not by proving theories but rather by falsifying them \cite{sep-popper}. Suppose someone does an experiment in a scenario described by Figure~\ref{fig:direction} that manipulates $X$ and finds no significant change in $Y$. This falsifies the causal model $X \to Y$, but does not falsify the predictive model. If the status quo in this field is to use predictive models, then this experiment may not result in any scientific progress. Other studies will continue to support the predictive model's accuracy, and the field can remain conflicted.

We have already discussed how causality involves more assumptions and conclusions. Any of these that can be tested provide another target for potential falsification. As a result, the scientist in our publishing story above decided not to use a causal model. Irrespective of their individual success in publishing, is this a good thing for science in general?

It is a misconception about science that the goal of designing a model or theory is to have the closest possible fit to reality in general, or to be immune to any criticism or disproof. A scientific theory can have great explanatory power and application potential even if it is known to be incorrect. For example, suppose someone builds an algorithm for autonomous vehicle control that uses classical mechanics. It would not be interesting or useful to point out that classical mechanics is ``wrong'' \emph{unless}, say, that algorithm will be used for rockets traveling at relativistic speeds.
 
Falsification does not need to be a blindly adversarial process reaching for any conclusion that can be disproved. It can be done selectively, with attention given to those parts of a model that are consequential in the intended application. As an example consider causal analysis of discrimination. The interested parties may only be disputing the strength of causal determination along one pathway in a causal graph, and are not interested in testing every other falsifiable prediction of the model.

Assumptions can enable or block scientific progress. They can block progress if they are never questioned, which seems unlikely given the incentives to publish novel results. They cannot enable progress if they are never considered. And that may be a real risk if the scientific and machine learning communities do not learn to tolerate more assumptions in some applications. \\

\begin{position}[Outsourcing responsibility]
A preference for predictive models is often a sign of avoiding responsibility.
\label{pos5}
\end{position}

If a work's value derives mainly from possible causal conclusions, but the author does not make transparent causal assumptions, they are leaving it to readers to reach the intended conclusion. 

Researchers sometimes avoid making transparent causal assumptions out of fear. They do not want to be responsible for more modeling choices because each choice opens up another potential criticism. They prefer whichever methods seem to work with the least number of choices, automatically, on the default settings. And this may not be all wrong, in fact, except that the same researchers often want to interpret the results causally. If they have taken a statistics course they may know to avoid using the word ``cause'' when talking about the results, but they might get away with suggesting or hinting at causality with weasel words like $X$ ``drives,'' ``predicts,'' or ``is linked to'' $Y$. The responsibility for making causal conclusions falls to their audience, who, due to their distance from the work, may be less familiar with the limitations.

Using causal models can be a more honest and transparent way to answer the questions we truly want to ask. Instead of pretending that we are only interested in prediction, we can admit that we want to change the world. We are almost always going to use predictions for some purpose that involves taking some actions to try to change some outcome. So we can and should use a causal model to elaborate how we think this is going to work. \\

\begin{position}[Relevance to machine learning]
The machine learning community is uniquely well-positioned to integrate causality, but also has a disciplinary bias against it.
\label{pos6}
\end{position}

The overlap between machine learning and causality is growing in absolute terms (see \citet{makhlouf2020survey, yao2021survey, scholkopf2022causality, kaddour2022causal} for some recent surveys). This is natural since applications with causal models usually involve learning functions from data. Machine learning has developed many powerful algorithms for estimating functions and learning about structure, and it is often only some additional assumptions required to transform any of these into explicitly causal methods.

However, there are reasons to be concerned about the prospects of causality in machine learning.

First, computer science as a discipline prefers general purpose algorithms, i.e. ones which are ``assumption free'' or ``model free.'' This makes sense for building robust systems, since an algorithm that relies on specific assumptions can break down if conditions change so that the assumptions no longer hold. There is also an economic logic: a technological service that is more automatic and scalable has more potential customers. Algorithms using causal models are less automatic and scalable because they require more context or task-specific input. Decisions about causal directions might require domain knowledge, for example.

Second, for many years, the ``secret sauce'' of machine learning's success \cite{donoho201750} has been the Common Task Framework: a competition between algorithms on a fixed, well-defined task \cite{liberman2010obituary}. These tasks are usually about predictive accuracy or a set of predictive benchmarks. And, usually, the SOTA algorithms that gain the most success and attention in the field are highly complex predictive models where ``the goal is not interpretability'' \cite{breiman2001statistical}. It makes sense from the viewpoint of constrained optimization that if we evaluate models only on predictive accuracy then any constraints--for example, belonging to a class of certain interpretable causal models--will likely result in worse evaluations.

On the plus side, pragmatism in computer science does sometimes make room for works that pose or reformulate problems differently. This may allow some exceptional works to use causal modeling without assuming the goal of learning a ``correct'' model \cite{janzing2023reinterpreting}.

\paragraph{Speculation.} Allow us this brief departure from more evidenced argument. We speculate that the high degree of competition in machine learning conferences--with their low acceptance rates and focus on SOTA performance in predictive tasks--exacerbates some of the challenges for causal methods we have outlined. In particular, when referees enter a review process that they know has a low acceptance rate, they may be looking for anything they can criticize, and the additional assumptions of a causal model are a prime target.


\section{Principles} 
\label{sec:principles}

\subsection{Scientific Pragmatism} 
\label{sec:sciprag}

Pragmatism is a philosophical tradition that emphasizes usefulness \cite{sep-pragmatism}. Continuing with this tradition, we advocate thinking of scientific theories--and models in general--as tools. Instead of asking if a model is right, correct, or accurate, we should ask if it is \emph{useful for a certain purpose}. Thus, the fitness of a theory or model depends on the goals of the people using it.

Recall our previous example of the autonomous vehicle control algorithm. If this system will only be used for automobiles on the surface of our planet, then classical mechanics is a good fit. Quantum mechanics has not been falsified under any conditions, but it would be a bad fit for the design of such an algorithm.

Causal models can be useful as tools for a variety of purposes. They can serve as a notation and language for reasoning and conveniently communicating about relationships between sets of variables \cite{imbens2020potential} and, hence, for interpretability and explainability \cite{blobaum2017estimation, sani2020explaining, frye2020asymmetric, heskes2020causal, zhao_causal_2021, loftus2023causal}.
Causality can be used to analyze algorithms for discrimination or fairness \cite{kilbertus2017avoiding, kusner_counterfactual_2017, nabi2018fair, zhang2018fairness, chiappa2019path, yang_causal_2021}.

Machine learning researchers are likely already comfortable with the idea of selecting the right tool for a task. It is important that they remember there are tasks other than prediction with SOTA accuracy. And in some application domains, the principle of scientific pragmatism supports our Positions~\ref{pos3} and~\ref{pos4} as an important counterweight to scientific perfectionism.

The greatest challenge for a scientific pragmatist is the danger of ``cargo cult science'' \cite{feynman1998cargo} or its relative ``cargo cult programming'' \cite{raymond1997jargon}. These terms describe failures that occur when people select tools or attempt to build them without understanding how they work, based only on appearances of success. In the classic example of a cargo cult as described by Feynman, people attempt to build an aircraft, for example, by fashioning objects with similar appearances. The problem is that such an aircraft, built without the knowledge and tools of engineering, cannot function. While the problem is obvious in this example, it can be more subtle in cargo cult science where a model or method may appear to function. The problem of cargo cult machine learning may be worse still because machine learning tasks are highly standardized and its tools automated. Students of machine learning can rapidly achieve the appearance of success without understanding the tools they are using, knowing about their limitations, or developing skills to judge or build new tools.

Pragmatism can serve as a descriptive theory about practices that are already standard. We simply advocate that this is done consciously and honestly, acknowledging our purposes. When we select tools for predictive accuracy we should not expect this to necessarily increase scientific knowledge if, for example, causal interpretability is sacrificed in the process.


\subsection{Value Pluralism}
\label{sec:pluralism}

When selecting tools and judging their usefulness, we may not be able to reduce this judgment to the application of a single value \cite{sep-value-pluralism}. We have a plurality of values that are important to us, and the relevance of these values may change depending on our purposes. The challenge increases when our values are in conflict. Readers may be thinking that multi-objective optimization or multi-task learning can simply ``solve'' this problem for us, but different values may be incommensurable or not quantifiable \cite{sep-value-theory}. For example, we may desire both predictive accuracy and interpretability, and we may not have any quantified measures of interpretability.

Value pluralism is in strong conflict with machine learning's ``secret sauce'' as described in Position~\ref{pos5}. The field has become more pluralistic by widening the set of standard tasks, and this is good work that should continue. However, there may still be limits on which values can be included for tasks that are standardized, or in systems built using automation and scaling \cite{NguyenForthcoming-NGUVCH, value_lockin}. Some values may require something closer to case-by-case human judgments. For example, the usefulness of a particular causal model may not be fully determined by the available data and algorithms, but also depend on a subjective judgment involving its interpretation.

\subsection{Humanism or Anthropocentrism}
\label{sec:humanism}

When we build infrastructure and systems using automation, selecting tools for their scalability and plug-and-play fitness within the existing system, it is not just causal models that are at risk of being pushed out of the picture. Another risk is evident in the terms ``human-in-the-loop'' or ``human feedback,'' the very existence of these terms implying that there are alternatives which somehow do not involve humans.

We assert that humans are the ultimate source and arbiters of values.\footnote{We are not opposed to a view that includes other forms of biological life, but we leave that for another discussion.} By bringing the focus to humans using tools, pragmatism could shed light on various perennially confusing issues in machine learning. It can help us remember that the ``agents'' are humans, and that software processes started by humans are tools being used for a purpose. 

In this human-centric view, it does not matter as much that causal methods may be less automatic or scalable. It does not matter as much when a particular algorithm is the SOTA for a certain task. We always have other purposes and values which are not codified in that task, and may not even be readily quantifiable. For example, we may sometimes wish to use a causal model within the context of a broad scientific question or business plan that involves various subjective factors. We may be motivated by those factors to make assumptions in that causal model which are not decidable in an automated way based on the available data. We can remember that this introduces limitations, and we can decide that the choice still fits with our values.

\section{Conclusion} 
\label{sec:conclusion}

\subsection{Related Ideas} 
\label{sec:context}

Inspired by some earlier debates in statistics, \citet{tukey1960conclusions} proposed we distinguish between two different goals or modes of statistical analysis: conclusions vs decisions. Conclusions (the Fisherian mode) should be final, authoritative, well-tested, and we should rarely need to revise them, if ever. As such, it makes sense to have high standards of evidence for conclusions. On the other hand, decisions (the Neyman-Pearson mode) are temporary, we often make them just so we can proceed to act, and we are not surprised if we need to try something else. We speculate that our Positions~\ref{pos2} and~\ref{pos3} are often related to a failure to make the distinction Tukey recommends here. In particular, we can make the decision to use a causal model without taking that model as conclusive.

Echoing through history, the pragmatic statistician George Box reminds us: ``All models are wrong, but some are useful,'' and ``Since all models are wrong the scientist must be alert to what is \emph{importantly wrong} [emphasis added]. It is inappropriate to be concerned about safety from mice when there are tigers abroad'' \cite{box1976science}.

Preceding the development of structural causal models with their formal definition of intervention, philosopher of science \citet{hacking1983representing} makes a case that the practice of science involves intervention rather than representation. This makes scientists pragmatic by necessity.

In the early days of machine learning, \citet{breiman2001statistical} argued in favor of its single-minded focus on predictive accuracy, and, more recently, \citet{rodu2023black} described conditions when such ``outcome reasoning'' may succeed or may be inappropriate.

Much recent work has focused on the problem of distribution shift or the ability of predictive models to generalize out-of-distribution. \citet{scholkopf2021toward} showed causal models can be a natural fit for this task. Interestingly, \citet{richens2024robust} established something like a converse of this result: if a predictive model can succeed under sufficiently many distribution shifts then that model entails an approximate causal model of the data generating process. This implies that if someone cares about maintaining predictive accuracy across a variety of different settings then, in a sense, they must care about causality. 


\subsection{Significance} 
\label{sec:significance}

Should anyone who does not personally have any uses for causal models care about challenges holding back the causal revolution? Yes, if they want humanity to continue unlocking the benefits of scientific progress, or if they want human values that are not in harmony with automated systems to be able to flourish.

A pragmatic use of causal models could revolutionize science. It has the potential to become a new standard language for the communication of scientific results, replacing the less interesting significance tests for differences between groups or for regression coefficients. For the sake of simplicity let's take the view that a scientific discovery occurs when someone falsifies a hypothesis. If that hypothesis is uninteresting, then it is not much of a discovery. Causal models are an ideal framework for specifying and communicating \emph{interesting} hypotheses. 

It is an empirical fact that humans have many different values. Systems using machine learning apparently grow most rapidly when they are built based on automation, scalability, and a narrow focus on standardized tasks. Such systems risk crowding out values which are not included in the set of tasks, or which are at odds with automation and scale.

\subsection{Summary} 
\label{sec:summary}

We have argued that methods based on causality face some challenges which can put them at a disadvantage. Within machine learning specifically, this occurs because of a narrow focus on SOTA performance on prediction tasks. Most of those tasks do not encode subjective human uses like causal interpretation, for example. We advocate a philosophy combining scientific pragmatism, value pluralism, and humanism, where models and tools can be selected based on a more inclusive set of values which may not be easily quantifiable or standardized. With these views, the machine learning community can make room for a greater variety of interesting applications and methods, including more that focus on causality, and promote the flourishing of more values aside from predictive accuracy. The causal revolution needs scientific pragmatism in order to proceed beyond its first steps, and human-centric value pluralism can help us apply scientific pragmatism wisely.

\section*{Acknowledgements}

The author would like to thank Pooja Loftus, Ricardo Silva, and David Watson for helpful conversations and feedback on this work.

\section*{Impact Statement}

This paper presents positions intended to advance the field of Machine Learning. There are many potential societal consequences, some of which were discussed throughout the text. We do not feel we should specifically highlight any other potential impacts here.

\nocite{langley00}

\bibliography{icml2024}

\begin{thebibliography}{53}
\providecommand{\natexlab}[1]{#1}
\providecommand{\url}[1]{\texttt{#1}}
\expandafter\ifx\csname urlstyle\endcsname\relax
  \providecommand{\doi}[1]{doi: #1}\else
  \providecommand{\doi}{doi: \begingroup \urlstyle{rm}\Url}\fi

\bibitem[Baker(2022)]{sep-simplicity}
Baker, A.
\newblock {Simplicity}.
\newblock In Zalta, E.~N. (ed.), \emph{The {Stanford} Encyclopedia of Philosophy}. Metaphysics Research Lab, Stanford University, {S}ummer 2022 edition, 2022.

\bibitem[Bl{\"o}baum \& Shimizu(2017)Bl{\"o}baum and Shimizu]{blobaum2017estimation}
Bl{\"o}baum, P. and Shimizu, S.
\newblock Estimation of interventional effects of features on prediction.
\newblock In \emph{2017 IEEE 27th International Workshop on Machine Learning for Signal Processing (MLSP)}, pp.\  1--6. IEEE, 2017.

\bibitem[Box(1976)]{box1976science}
Box, G.~E.
\newblock Science and statistics.
\newblock \emph{Journal of the American Statistical Association}, 71\penalty0 (356):\penalty0 791--799, 1976.

\bibitem[Breiman(2001)]{breiman2001statistical}
Breiman, L.
\newblock Statistical modeling: The two cultures (with comments and a rejoinder by the author).
\newblock \emph{Statistical science}, 16\penalty0 (3):\penalty0 199--231, 2001.

\bibitem[Bynum et~al.(2021)Bynum, Loftus, and Stoyanovich]{bynum_disaggregated_2021}
Bynum, L., Loftus, J., and Stoyanovich, J.
\newblock Disaggregated {Interventions} to {Reduce} {Inequality}.
\newblock In \emph{Equity and {Access} in {Algorithms}, {Mechanisms}, and {Optimization}}, pp.\  1--13. Association for Computing Machinery, New York, NY, USA, October 2021.
\newblock ISBN 978-1-4503-8553-4.
\newblock URL \url{https://doi.org/10.1145/3465416.3483286}.

\bibitem[Bynum et~al.(2023)Bynum, Loftus, and Stoyanovich]{bynum2023counterfactuals}
Bynum, L., Loftus, J., and Stoyanovich, J.
\newblock Counterfactuals for the future.
\newblock In \emph{Proceedings of the AAAI Conference on Artificial Intelligence}, 2023.

\bibitem[Bynum et~al.(2024)Bynum, Loftus, and Stoyanovich]{bynum2024new}
Bynum, L.~E., Loftus, J.~R., and Stoyanovich, J.
\newblock A new paradigm for counterfactual reasoning in fairness and recourse.
\newblock In \emph{Proceedings of the Thirty-Third International Joint Conference on Artificial Intelligence (IJCAI)}, 2024.

\bibitem[Chiappa(2019)]{chiappa2019path}
Chiappa, S.
\newblock Path-specific counterfactual fairness.
\newblock In \emph{Proceedings of the AAAI Conference on Artificial Intelligence}, volume~33, pp.\  7801--7808, 2019.

\bibitem[Cofield et~al.(2010)Cofield, Corona, and Allison]{cofield2010use}
Cofield, S.~S., Corona, R.~V., and Allison, D.~B.
\newblock Use of causal language in observational studies of obesity and nutrition.
\newblock \emph{Obesity facts}, 3\penalty0 (6):\penalty0 353--356, 2010.

\bibitem[Donoho(2017)]{donoho201750}
Donoho, D.
\newblock 50 years of data science.
\newblock \emph{Journal of Computational and Graphical Statistics}, 26\penalty0 (4):\penalty0 745--766, 2017.

\bibitem[Feynman(1998)]{feynman1998cargo}
Feynman, R.~P.
\newblock Cargo cult science.
\newblock In \emph{The art and science of analog circuit design}, pp.\  55--61. Elsevier, 1998.

\bibitem[Finnveden et~al.(2022)Finnveden, Riedel, and Shulman]{value_lockin}
Finnveden, L., Riedel, J., and Shulman, C.
\newblock {AGI and Lock-In}.
\newblock \url{https://forum.effectivealtruism.org/posts/KqCybin8rtfP3qztq/agi-and-lock-in}, 2022.

\bibitem[Frye et~al.(2020)Frye, Rowat, and Feige]{frye2020asymmetric}
Frye, C., Rowat, C., and Feige, I.
\newblock Asymmetric shapley values: incorporating causal knowledge into model-agnostic explainability.
\newblock \emph{Advances in Neural Information Processing Systems}, 33:\penalty0 1229--1239, 2020.

\bibitem[Haber et~al.(2018)Haber, Smith, Moscoe, Andrews, Audy, Bell, Brennan, Breskin, Kane, Karra, et~al.]{haber2018causal}
Haber, N., Smith, E.~R., Moscoe, E., Andrews, K., Audy, R., Bell, W., Brennan, A.~T., Breskin, A., Kane, J.~C., Karra, M., et~al.
\newblock Causal language and strength of inference in academic and media articles shared in social media (claims): A systematic review.
\newblock \emph{PloS one}, 13\penalty0 (5):\penalty0 e0196346, 2018.

\bibitem[Haber et~al.(2022)Haber, Wieten, Rohrer, Arah, Tennant, Stuart, Murray, Pilleron, Lam, Riederer, et~al.]{haber2022causal}
Haber, N.~A., Wieten, S.~E., Rohrer, J.~M., Arah, O.~A., Tennant, P.~W., Stuart, E.~A., Murray, E.~J., Pilleron, S., Lam, S.~T., Riederer, E., et~al.
\newblock Causal and associational language in observational health research: a systematic evaluation.
\newblock \emph{American journal of epidemiology}, 191\penalty0 (12):\penalty0 2084--2097, 2022.

\bibitem[Hacking(1983)]{hacking1983representing}
Hacking, I.
\newblock \emph{Representing and intervening: Introductory topics in the philosophy of natural science}.
\newblock Cambridge university press, 1983.

\bibitem[Han et~al.(2022)Han, Leung, Storman, Xiao, Srivastava, Talukdar, El~Dib, Morassut, Zeraatkar, Johnston, et~al.]{han2022causal}
Han, M.~A., Leung, G., Storman, D., Xiao, Y., Srivastava, A., Talukdar, J.~R., El~Dib, R., Morassut, R.~E., Zeraatkar, D., Johnston, B.~C., et~al.
\newblock Causal language use in systematic reviews of observational studies is often inconsistent with intent: a systematic survey.
\newblock \emph{Journal of Clinical Epidemiology}, 148:\penalty0 65--73, 2022.

\bibitem[Heskes et~al.(2020)Heskes, Sijben, Bucur, and Claassen]{heskes2020causal}
Heskes, T., Sijben, E., Bucur, I.~G., and Claassen, T.
\newblock Causal shapley values: Exploiting causal knowledge to explain individual predictions of complex models.
\newblock \emph{Advances in neural information processing systems}, 33:\penalty0 4778--4789, 2020.

\bibitem[Hu(2023)]{hu2023race}
Hu, L.
\newblock What is “race” in algorithmic discrimination on the basis of race?
\newblock \emph{Journal of Moral Philosophy}, 1\penalty0 (aop):\penalty0 1--26, 2023.

\bibitem[Hu \& Kohler-Hausmann(2020)Hu and Kohler-Hausmann]{hu2020s}
Hu, L. and Kohler-Hausmann, I.
\newblock What's sex got to do with machine learning?
\newblock In \emph{Proceedings of the 2020 Conference on Fairness, Accountability, and Transparency}, pp.\  513--513, 2020.

\bibitem[Imbens(2020)]{imbens2020potential}
Imbens, G.~W.
\newblock Potential outcome and directed acyclic graph approaches to causality: Relevance for empirical practice in economics.
\newblock \emph{Journal of Economic Literature}, 58\penalty0 (4):\penalty0 1129--1179, 2020.

\bibitem[Jacobs \& Wallach(2021)Jacobs and Wallach]{jacobs2021measurement}
Jacobs, A.~Z. and Wallach, H.
\newblock Measurement and fairness.
\newblock In \emph{Proceedings of the 2021 ACM conference on fairness, accountability, and transparency}, pp.\  375--385, 2021.

\bibitem[Janzing et~al.(2023)Janzing, Faller, and Vankadara]{janzing2023reinterpreting}
Janzing, D., Faller, P.~M., and Vankadara, L.~C.
\newblock Reinterpreting causal discovery as the task of predicting unobserved joint statistics.
\newblock \emph{arXiv preprint arXiv:2305.06894}, 2023.

\bibitem[Kaddour et~al.(2022)Kaddour, Lynch, Liu, Kusner, and Silva]{kaddour2022causal}
Kaddour, J., Lynch, A., Liu, Q., Kusner, M.~J., and Silva, R.
\newblock Causal machine learning: A survey and open problems.
\newblock \emph{arXiv preprint arXiv:2206.15475}, 2022.

\bibitem[Kilbertus et~al.(2017)Kilbertus, Rojas~Carulla, Parascandolo, Hardt, Janzing, and Sch{\"o}lkopf]{kilbertus2017avoiding}
Kilbertus, N., Rojas~Carulla, M., Parascandolo, G., Hardt, M., Janzing, D., and Sch{\"o}lkopf, B.
\newblock Avoiding discrimination through causal reasoning.
\newblock \emph{Advances in neural information processing systems}, 30, 2017.

\bibitem[Kohler-Hausmann(2018)]{kohler2018eddie}
Kohler-Hausmann, I.
\newblock {Eddie Murphy and the dangers of counterfactual causal thinking about detecting racial discrimination}.
\newblock \emph{Nw. UL Rev.}, 113:\penalty0 1163, 2018.

\bibitem[Kusner et~al.(2017)Kusner, Loftus, Russell, and Silva]{kusner_counterfactual_2017}
Kusner, M.~J., Loftus, J., Russell, C., and Silva, R.
\newblock Counterfactual {Fairness}.
\newblock In \emph{Advances in {Neural} {Information} {Processing} {Systems}}, volume~30. Curran Associates, Inc., 2017.
\newblock URL \url{https://proceedings.neurips.cc/paper/2017/hash/a486cd07e4ac3d270571622f4f316ec5-Abstract.html}.

\bibitem[Langley(2000)]{langley00}
Langley, P.
\newblock Crafting papers on machine learning.
\newblock In Langley, P. (ed.), \emph{Proceedings of the 17th International Conference on Machine Learning (ICML 2000)}, pp.\  1207--1216, Stanford, CA, 2000. Morgan Kaufmann.

\bibitem[Lazarus et~al.(2015)Lazarus, Haneef, Ravaud, and Boutron]{lazarus2015classification}
Lazarus, C., Haneef, R., Ravaud, P., and Boutron, I.
\newblock Classification and prevalence of spin in abstracts of non-randomized studies evaluating an intervention.
\newblock \emph{BMC medical research methodology}, 15:\penalty0 1--8, 2015.

\bibitem[Legg \& Hookway(2021)Legg and Hookway]{sep-pragmatism}
Legg, C. and Hookway, C.
\newblock {Pragmatism}.
\newblock In Zalta, E.~N. (ed.), \emph{The {Stanford} Encyclopedia of Philosophy}. Metaphysics Research Lab, Stanford University, {S}ummer 2021 edition, 2021.

\bibitem[Liberman(2010)]{liberman2010obituary}
Liberman, M.
\newblock Obituary: Fred jelinek.
\newblock \emph{Computational Linguistics}, 36\penalty0 (4):\penalty0 595--599, 2010.

\bibitem[Loftus et~al.(2023)Loftus, Bynum, and Hansen]{loftus2023causal}
Loftus, J.~R., Bynum, L.~E., and Hansen, S.
\newblock Causal dependence plots.
\newblock \emph{arXiv preprint arXiv:2303.04209}, 2023.

\bibitem[Makhlouf et~al.(2020)Makhlouf, Zhioua, and Palamidessi]{makhlouf2020survey}
Makhlouf, K., Zhioua, S., and Palamidessi, C.
\newblock Survey on causal-based machine learning fairness notions.
\newblock \emph{arXiv preprint arXiv:2010.09553}, 2020.

\bibitem[Mason(2023)]{sep-value-pluralism}
Mason, E.
\newblock {Value Pluralism}.
\newblock In Zalta, E.~N. and Nodelman, U. (eds.), \emph{The {Stanford} Encyclopedia of Philosophy}. Metaphysics Research Lab, Stanford University, {S}ummer 2023 edition, 2023.

\bibitem[Nabi \& Shpitser(2018)Nabi and Shpitser]{nabi2018fair}
Nabi, R. and Shpitser, I.
\newblock Fair inference on outcomes.
\newblock In \emph{Proceedings of the AAAI Conference on Artificial Intelligence}, volume~32, 2018.

\bibitem[Nguyen(forthcoming)]{NguyenForthcoming-NGUVCH}
Nguyen, C.~T.
\newblock Value capture.
\newblock \emph{Journal of Ethics and Social Philosophy}, forthcoming.

\bibitem[{Open Science Collaboration}(2015)]{open2015estimating}
{Open Science Collaboration}.
\newblock Estimating the reproducibility of psychological science.
\newblock \emph{Science}, 349\penalty0 (6251):\penalty0 aac4716, 2015.

\bibitem[Prasad et~al.(2013)Prasad, Jorgenson, Ioannidis, and Cifu]{prasad2013observational}
Prasad, V., Jorgenson, J., Ioannidis, J.~P., and Cifu, A.
\newblock Observational studies often make clinical practice recommendations: an empirical evaluation of authors' attitudes.
\newblock \emph{Journal of clinical epidemiology}, 66\penalty0 (4):\penalty0 361--366, 2013.

\bibitem[Raymond \& Steele(1997)Raymond and Steele]{raymond1997jargon}
Raymond, E.~S. and Steele, G.~L.
\newblock \emph{The Jargon File, Version 4.0. 0, 24 Jul 1996}.
\newblock Project Gutenberg, 1997.

\bibitem[Richens \& Everitt(2024)Richens and Everitt]{richens2024robust}
Richens, J. and Everitt, T.
\newblock Robust agents learn causal world models.
\newblock \emph{International Conference on Learning Representations}, 2024.

\bibitem[Rodu \& Baiocchi(2023)Rodu and Baiocchi]{rodu2023black}
Rodu, J. and Baiocchi, M.
\newblock When black box algorithms are (not) appropriate.
\newblock \emph{Observational Studies}, 9\penalty0 (2):\penalty0 79--101, 2023.

\bibitem[Sani et~al.(2020)Sani, Malinsky, and Shpitser]{sani2020explaining}
Sani, N., Malinsky, D., and Shpitser, I.
\newblock Explaining the behavior of black-box prediction algorithms with causal learning.
\newblock \emph{arXiv preprint arXiv:2006.02482}, 2020.

\bibitem[Sch{\"o}lkopf(2022)]{scholkopf2022causality}
Sch{\"o}lkopf, B.
\newblock Causality for machine learning.
\newblock In \emph{Probabilistic and Causal Inference: The Works of Judea Pearl}, pp.\  765--804. 2022.

\bibitem[Sch{\"o}lkopf et~al.(2021)Sch{\"o}lkopf, Locatello, Bauer, Ke, Kalchbrenner, Goyal, and Bengio]{scholkopf2021toward}
Sch{\"o}lkopf, B., Locatello, F., Bauer, S., Ke, N.~R., Kalchbrenner, N., Goyal, A., and Bengio, Y.
\newblock Toward causal representation learning.
\newblock \emph{Proceedings of the IEEE}, 109\penalty0 (5):\penalty0 612--634, 2021.

\bibitem[Schroeder(2021)]{sep-value-theory}
Schroeder, M.
\newblock {Value Theory}.
\newblock In Zalta, E.~N. (ed.), \emph{The {Stanford} Encyclopedia of Philosophy}. Metaphysics Research Lab, Stanford University, {F}all 2021 edition, 2021.

\bibitem[Sen \& Wasow(2016)Sen and Wasow]{sen2016race}
Sen, M. and Wasow, O.
\newblock Race as a bundle of sticks: Designs that estimate effects of seemingly immutable characteristics.
\newblock \emph{Annual Review of Political Science}, 19:\penalty0 499--522, 2016.

\bibitem[Thornton(2023)]{sep-popper}
Thornton, S.
\newblock {Karl Popper}.
\newblock In Zalta, E.~N. and Nodelman, U. (eds.), \emph{The {Stanford} Encyclopedia of Philosophy}. Metaphysics Research Lab, Stanford University, {W}inter 2023 edition, 2023.

\bibitem[Tukey(1960)]{tukey1960conclusions}
Tukey, J.~W.
\newblock Conclusions vs decisions.
\newblock \emph{Technometrics}, 2\penalty0 (4):\penalty0 423--433, 1960.

\bibitem[Yang et~al.(2021)Yang, Loftus, and Stoyanovich]{yang_causal_2021}
Yang, K., Loftus, J.~R., and Stoyanovich, J.
\newblock Causal {Intersectionality} and {Fair} {Ranking}.
\newblock In Ligett, K. and Gupta, S. (eds.), \emph{2nd {Symposium} on {Foundations} of {Responsible} {Computing} ({FORC} 2021)}, volume 192 of \emph{Leibniz {International} {Proceedings} in {Informatics} ({LIPIcs})}, pp.\  7:1--7:20, Dagstuhl, Germany, 2021. Schloss Dagstuhl – Leibniz-Zentrum für Informatik.
\newblock ISBN 978-3-95977-187-0.
\newblock \doi{10.4230/LIPIcs.FORC.2021.7}.
\newblock URL \url{https://drops.dagstuhl.de/opus/volltexte/2021/13875}.
\newblock ISSN: 1868-8969.

\bibitem[Yao et~al.(2021)Yao, Chu, Li, Li, Gao, and Zhang]{yao2021survey}
Yao, L., Chu, Z., Li, S., Li, Y., Gao, J., and Zhang, A.
\newblock A survey on causal inference.
\newblock \emph{ACM Transactions on Knowledge Discovery from Data (TKDD)}, 15\penalty0 (5):\penalty0 1--46, 2021.

\bibitem[Yu et~al.(2019)Yu, Li, and Wang]{yu2019detecting}
Yu, B., Li, Y., and Wang, J.
\newblock Detecting causal language use in science findings.
\newblock In \emph{Proceedings of the 2019 Conference on Empirical Methods in Natural Language Processing and the 9th International Joint Conference on Natural Language Processing (EMNLP-IJCNLP)}, pp.\  4664--4674, 2019.

\bibitem[Zhang \& Bareinboim(2018)Zhang and Bareinboim]{zhang2018fairness}
Zhang, J. and Bareinboim, E.
\newblock Fairness in decision-making—the causal explanation formula.
\newblock In \emph{Proceedings of the AAAI Conference on Artificial Intelligence}, volume~32, 2018.

\bibitem[Zhao \& Hastie(2021)Zhao and Hastie]{zhao_causal_2021}
Zhao, Q. and Hastie, T.
\newblock Causal interpretations of black-box models.
\newblock \emph{Journal of Business \& Economic Statistics}, 39\penalty0 (1):\penalty0 272--281, 2021.
\newblock \doi{10.1080/07350015.2019.1624293}.
\newblock URL \url{https://doi.org/10.1080/07350015.2019.1624293}.

\end{thebibliography}
\bibliographystyle{icml2024}

\end{document}